\documentclass[manuscript]{acmart}
\usepackage{subcaption}
\usepackage{graphicx}
\AtBeginDocument{%
  \providecommand\BibTeX{{%
    \normalfont B\kern-0.5em{\scshape i\kern-0.25em b}\kern-0.8em\TeX}}}

\copyrightyear{2023} 
\acmYear{2023} 
\setcopyright{rightsretained} 
\acmConference[MobileHCI '23 Companion]{25th International Conference on Mobile Human-Computer Interaction}{September 26--29, 2023}{Athens, Greece}
\acmBooktitle{25th International Conference on Mobile Human-Computer Interaction (MobileHCI '23 Companion), September 26--29, 2023, Athens, Greece}
\acmDOI{10.1145/3565066.3608698}
\acmISBN{978-1-4503-9924-1/23/09}

\begin{document}

\title[Decoding Emotional Valence from Wearables]{Decoding Emotional Valence from Wearables: Can Our Data Reveal Our True Feelings?}

\author{Michal K. Grzeszczyk}
\email{{m.grzeszczyk@sanoscience.org}}
\affiliation{%
  \institution{Sano Centre for Computational Medicine}
  \city{Cracow}
  \country{Poland}
}

\author{Anna Lisowska}
\affiliation{%
  \institution{Poznan University of Technology}
  \city{Poznan}
  \country{Poland}
}

\author{Arkadiusz Sitek}
\affiliation{%
  \institution{Massachusetts General Hospital, Harvard Medical School}
  \city{Boston}
  \country{USA}
}

\author{Aneta Lisowska}
\affiliation{%
  \institution{Poznan University of Technology}
  \city{Poznan}
  \country{Poland}
}

\renewcommand{\shortauthors}{Michal K. Grzeszczyk et al.}

\begin{abstract}
Automatic detection and tracking of emotional states has the potential for helping individuals with various mental health conditions. While previous studies have captured physiological signals using wearable devices in laboratory settings, providing valuable insights into the relationship between physiological responses and mental states, the transfer of these findings to real-life scenarios is still in its nascent stages.
Our research aims to bridge the gap between laboratory-based studies and real-life settings by leveraging consumer-grade wearables and self-report measures. We conducted a preliminary study involving 15 healthy participants to assess the efficacy of wearables in capturing user valence in real-world settings. In this paper, we present the initial analysis of the collected data, focusing primarily on the results of valence classification. Our findings demonstrate promising results in distinguishing between high and low positive valence, achieving an F1 score of 0.65. This research opens up avenues for future research in the field of mobile mental health interventions.
\end{abstract}





\maketitle

\section{Introduction}

Maintaining emotional well-being is important for overall health and can significantly impact an individual's quality of life. Automatic detection and tracking of emotional states could be beneficial for individuals with various mental health conditions, helping them better understand their symptoms and manage their emotional well-being \cite{patel2018apps}. This could potentially lead to earlier interventions and more personalized treatment plans tailored to the individual's needs \cite{patoz2021patient}.

However, current mobile well-being interventions often rely on self-reported data from users to monitor their emotional states. While this approach can be useful, it can be burdensome for the user, and the data obtained may lack consistency \cite{oakley2021works}. An alternative approach is to pair mobile applications with wearable devices for automated tracking of physiological states, which can reduce the effort required from users and provide more objective and consistent data.

Efforts have been made to detect emotional states using physiological signals captured by wearable devices. Significant progress has been made in laboratory conditions \cite{schmidt2018introducing,markova2019clas} and settings designed to resemble natural conditions \cite{park2020k}. However, challenges still exist for detecting emotional states in real-world scenarios. For example, many consumer-grade devices offer automatic stress tracking \cite{peake2018critical}, but they do not take into account the valence of the emotional state, which is essential for distinguishing between positive stress, such as excitement, and negative stress.

We conducted a preliminary study with a cohort of 15 healthy participants to assess the efficacy of consumer-grade wearables in capturing user valence in real-world settings. Throughout a two-week period, participants provided multiple self-reported emotional state ratings each day, while their physiological signals were concurrently recorded by the wearable devices (see Section \ref{sec:method}).
In this paper, we present the initial analysis of the collected data, focusing primarily on the results of valence classification (see Section \ref{sec:results}). The preliminary findings in distinguishing between high and low positive valence in real-life settings are promising, achieving an F1 score of 0.65. However, further investigations are required to determine if the classification results can be further improved by leveraging physiological signals alone or if the inclusion of additional contextual features is necessary. Furthermore, we discuss the implications of our findings for future studies that employ mobile and wearable devices for the collection of emotional and mental state data (see Section \ref{sec:discussion}). These implications encompass considerations related to the development of more accurate and robust classification models and the potential application of our research in various domains such as mental health and well-being.

\section{Related Work}
Datasets such as WESAD (Wearable Stress and Affect Detection) \cite{schmidt2018introducing}, CLAS (Cognitive Load, Affect and Stress Recognition)\cite{markova2019clas}, K-EmoCon \cite{park2020k} and CogWear \cite{grzeszczykcogwear} have played a crucial role in the development of machine learning models for detecting mental states, including amusement, stress, and high cognitive load. These datasets have captured physiological signals using wearable devices in laboratory settings, providing valuable insights into the relationship between physiological responses and mental states. However, there is a growing recognition that investigating emotions in laboratory environments may not fully capture the complexity of these phenomena in real-life contexts \cite{picard2016automating}.

To address this limitation, researchers utilise consumer-grade wearables, which have emerged as a powerful tool for capturing emotions in everyday life. These devices offer a convenient and non-intrusive means of continuously monitoring physiological signals associated with emotional experiences. In particular, the incorporation of photoplethysmography (PPG) sensors in consumer-grade wearables \cite{saganowski2020consumer} allows for the capture of the blood volume pulse signal. By extracting heart rate variability (HRV) features from the PPG signal, researchers can gain insights into an individual's mental and emotional state. For example, Can et al. \cite{can2019continuous} conducted a study focused on detecting high levels of stress in real-life settings. They utilized HRV features from the Empatica E4 wearable device, achieving a 90\% accuracy rate. Additionally, they extracted HRV features from signals captured by Samsung Gear S, resulting in an 84\% accuracy rate.

While stress detection from wearables has been a common research focus \cite{can2019stress}, it is equally important to differentiate between positive and negative valence to obtain a comprehensive understanding of emotional states. Therefore in this study, we consider positive and negative affect. Our research aims to bridge the gap between laboratory-based studies and real-life settings by leveraging consumer-grade wearables and self-report measures. We seek to capture emotions and cognitive load in individuals' natural environments, facilitating a deeper examination of the intricate relationship between physiological indicators, subjective experiences, and the differentiation of positive and negative valence.

\section{Method}
\label{sec:method}
In this feasibility study, we aimed to infer self-reported emotional state from the signal obtained from consumer-grade smartwatches. In what follows, we describe the data collection method. We present the application used for data acquisition and data pre-processing pipeline.

\subsection{Study Protocol}

A cohort of fifteen healthy volunteers, comprising ten females and five males aged between 26 and 55, were recruited for this study. The participants were equipped with Samsung Galaxy Watch 4 (SGW4) watches capable of capturing their blood volume pulse, enabling the extraction of heart rate variability features, as outlined in Section \ref{procesing}. To complement the physiological data, participants were prompted through notifications to engage in self-reporting of their emotional state at least twice daily. This comprehensive self-reporting encompassed the assessment of cognitive load using a 5-point Likert scale, reflecting the mental effort expended while performing the preceding task. Additionally, participants completed the PANAS-10 questionnaire \cite{thompson2007development}, a validated instrument encompassing both positive emotions (Active, Inspired, Attentive, Determined, Alert) and negative emotions (Hostile, Nervous, Upset, Afraid, Ashamed). To encourage self-reporting participants could unlock a well-being tip after each completed self-assessment. The study was approved by the AGH University of Science and Technology Ethics Committee (IRB number 3/2022).

\subsection{Wearable and Mobile App for Data Acquisition}

\begin{figure}[t]
    \centering
    \includegraphics[width=0.6\textwidth]{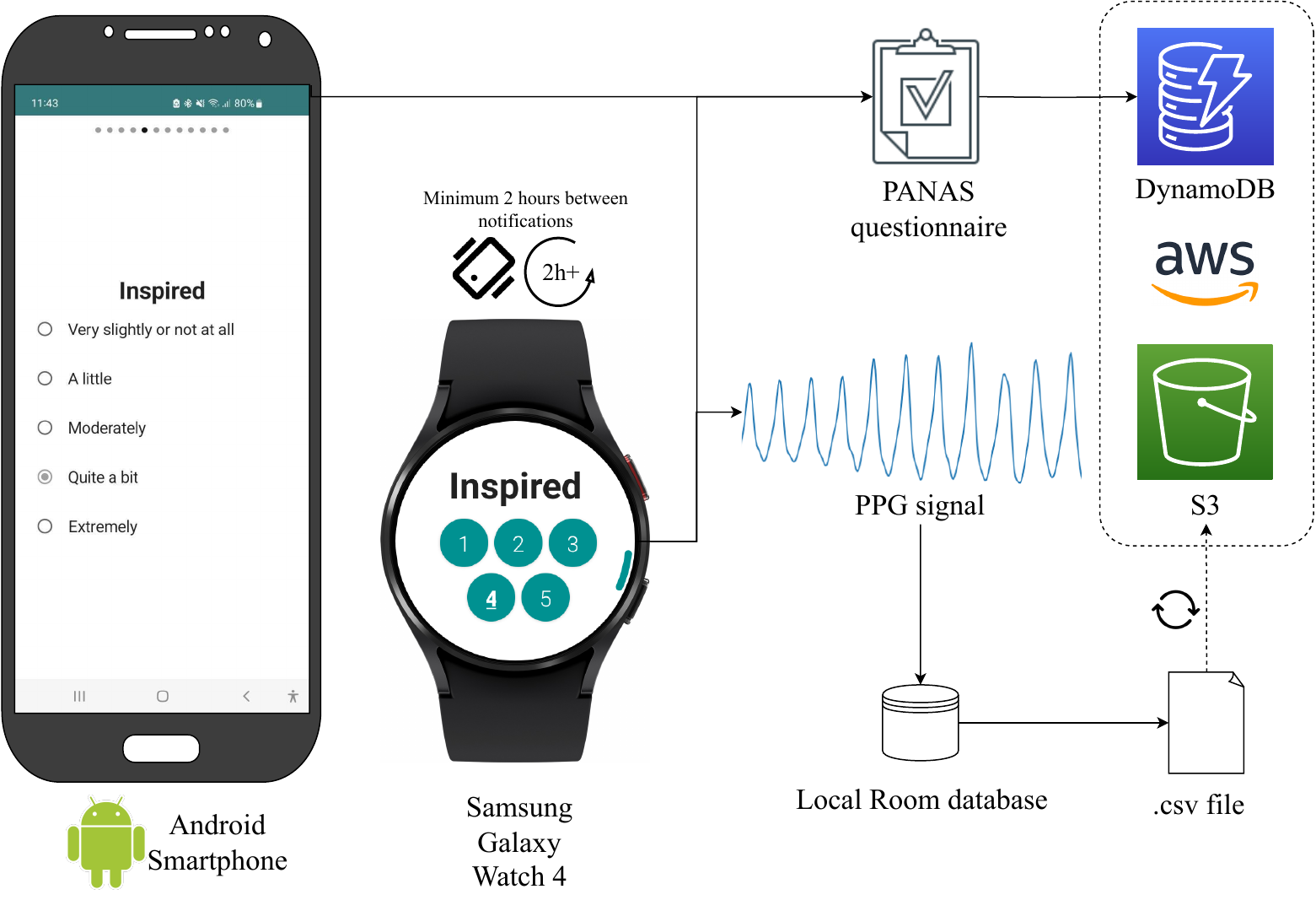}
    \caption{An overview of our data acquisition flow. The participants could fill in the PANAS questionnaire either on the Android smartphone or Samsung Galaxy Watch 4 (preferred option). The completed surveys were sent to the DynamoDB database. During the survey completion on the smartwatch, the PPG signal was collected and stored in the local database. The signal was then transformed and saved in .csv files which were later uploaded to the S3 bucket.}
    \label{fig:diagram}
\end{figure}

We developed a mobile application specifically designed for data acquisition purposes, which was implemented using Kotlin 1.6.10 as the code base and Jetpack Compose 1.5.0 for user interface development. The application was designed to be executed on two types of devices, namely Android smartphones and SGW4 smartwatches. The PANAS questionnaire could be completed on both devices. Since the PPG signal can be collected on SGW4 only, the participants were encouraged to use the smartwatch version of the app when possible. During working hours (10 a.m. - 6 p.m.), at most every 120 minutes, the notification on the smartwatch would remind participants to fill out the survey.

An overview of our data acquisition flow is presented in Fig. \ref{fig:diagram}. The questionnaire consisted of ten questions (plus a question about cognitive load), and participants were presented with only one question at a time on the screen. After answering the question, the next one was automatically displayed. Participants were able to navigate back and forth between previously completed questions by scrolling left or right. Upon completing the survey, the results were automatically sent to DynamoDB - a NoSQL database on the Amazon Web Services (AWS) cloud. 

If participants were using the smartwatch version of the application, during the survey answering process, the PPG Green signal was collected. The signal was acquired with Samsung Privilidged Health SDK (1.1.0) with 300 data points collected every 12 seconds (25 Hz frequency). To avoid loss of data samples transferred to the cloud and overload of the smartwatch during application usage, the signal was stored in the local Room database (2.4.2). Each survey was associated with a signal collected during its completion via a unique id. The PPG signal was then extracted from the local database into .csv files for each session, which were asynchronously uploaded to AWS Simple Storage Service (S3).

\subsection{Data Pre-Processing}
\label{procesing}

In order to align the physiological responses associated with the self-reported emotional states, we extracted raw PPG signals within a time window spanning 30 seconds before and after the reported emotional state.  To ensure data quality, we applied a bandpass filter implementation from the heartpy package \cite{van2018heart} to the extracted PPG signals. The primary objective of this filtering process was to eliminate noise and artifacts while retaining the frequency components essential for heart rate analysis. By utilizing the bandpass filter, we obtained a filtered PPG signal that served as the basis for extracting the following features:  \textit{BPM} (Beats per Minute), \textit{IBI} (Interbeat Interval), \textit{SDNN} (Standard Deviation of NN Intervals), \textit{RMSSD} (Root Mean Square of Successive Differences), the proportion of adjacent IBIs that differ by more than 20 milliseconds (\textit{pNN20}) or 50 milliseconds (\textit{pNN50}), HR-MAD (Heart Rate Mean Absolute Deviation), \textit{SD1} and \textit{SD2} - components of a geometric representation of HRV known as a Poincaré plot, \textit{S} (another component of the Poincaré plot, which represents the length of the line connecting adjacent NN intervals), \textit{SD1/SD2}, \textit{Breathing Rate}.

\section{Data Analysis}
\label{sec:results}
% We had 12 participants with features extracted from the blood volume signal who have given at least one response to the PANAS questionnaire at the time of the measurement. 

Overall, we had 482 responses. For each response, the scores for positive emotions were summed together to obtain an overall positive affect and the scores for negative emotions were summed together to obtain the overall negative affect (both in the range from 5 to 25).

\begin{figure}[t]
  \centering

  \begin{subfigure}[b]{0.232\textwidth}
    \centering
    \includegraphics[width=\textwidth]{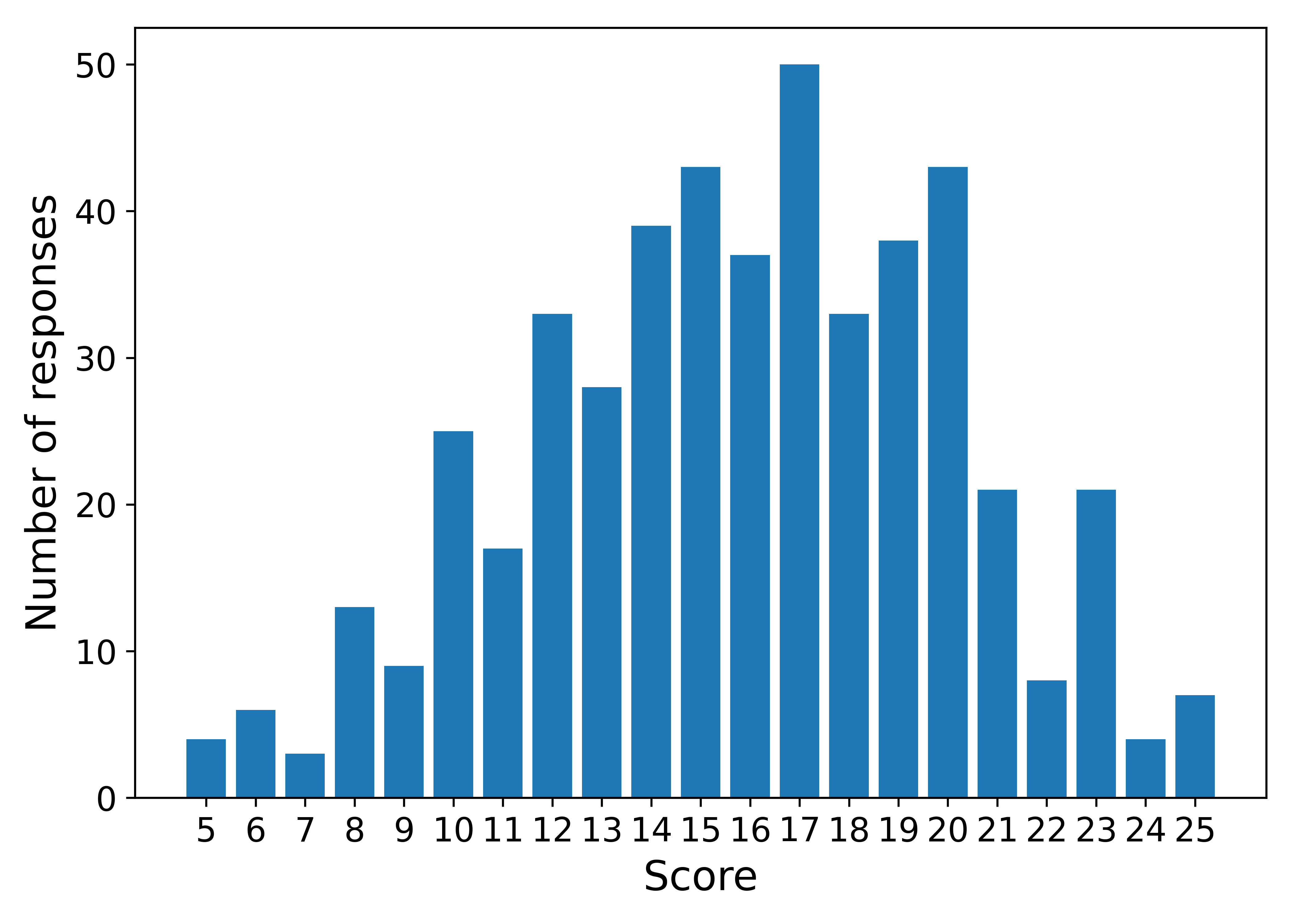}
    \caption{Positive Affect}
    \label{fig:paffect}
  \end{subfigure}
  \begin{subfigure}[b]{0.232\textwidth}
    \centering
    \includegraphics[width=\textwidth]{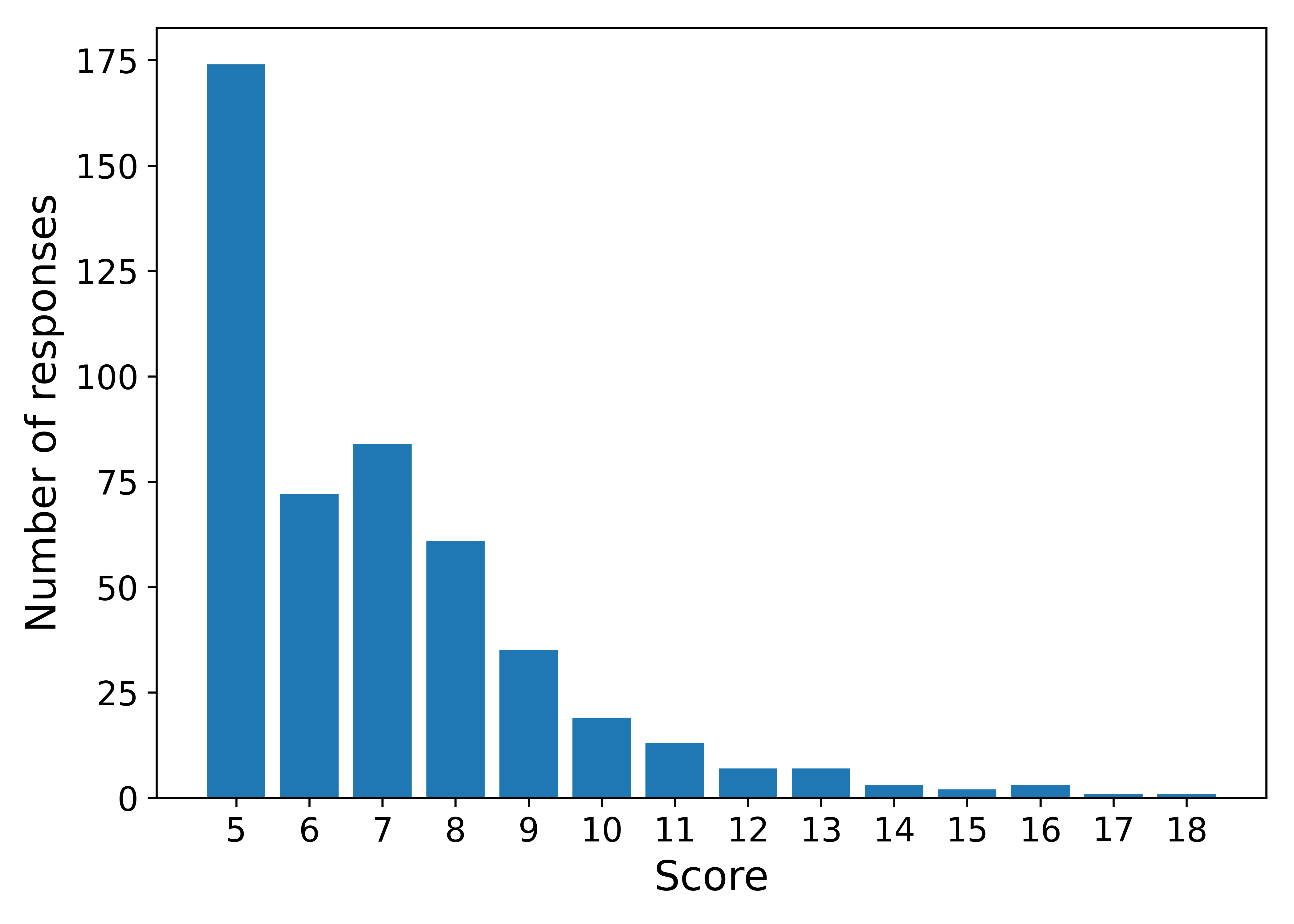}
    \caption{Negative Affect}
    \label{fig:naffect}
  \end{subfigure}
  \begin{subfigure}[b]{0.232\textwidth}
    \centering
    \includegraphics[width=\textwidth]{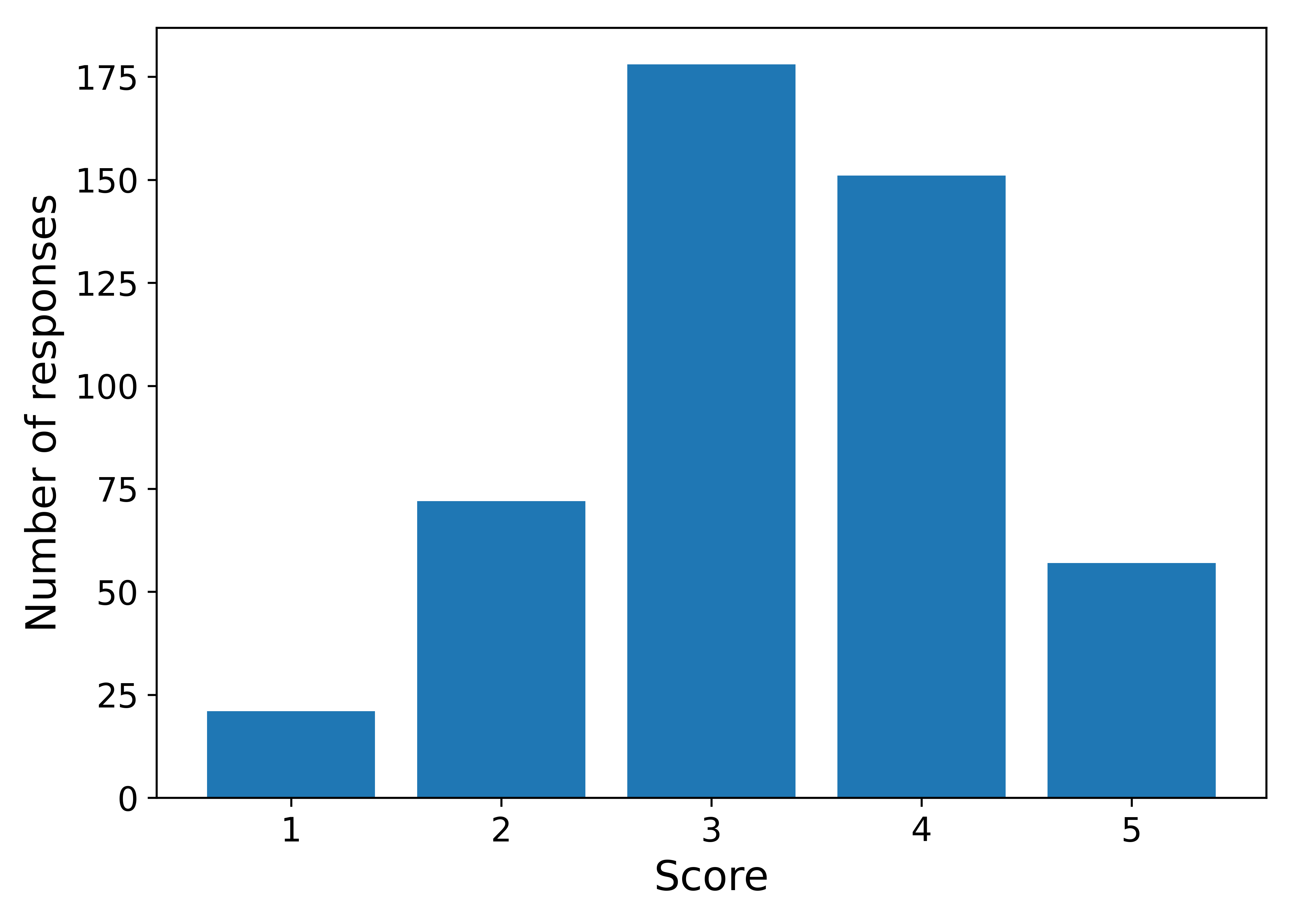}
    \caption{Cognitive Load}
    \label{fig:cognitive_load}
     \end{subfigure}
  \begin{subfigure}[b]{0.284\textwidth}
    \centering
    \includegraphics[width=\textwidth]{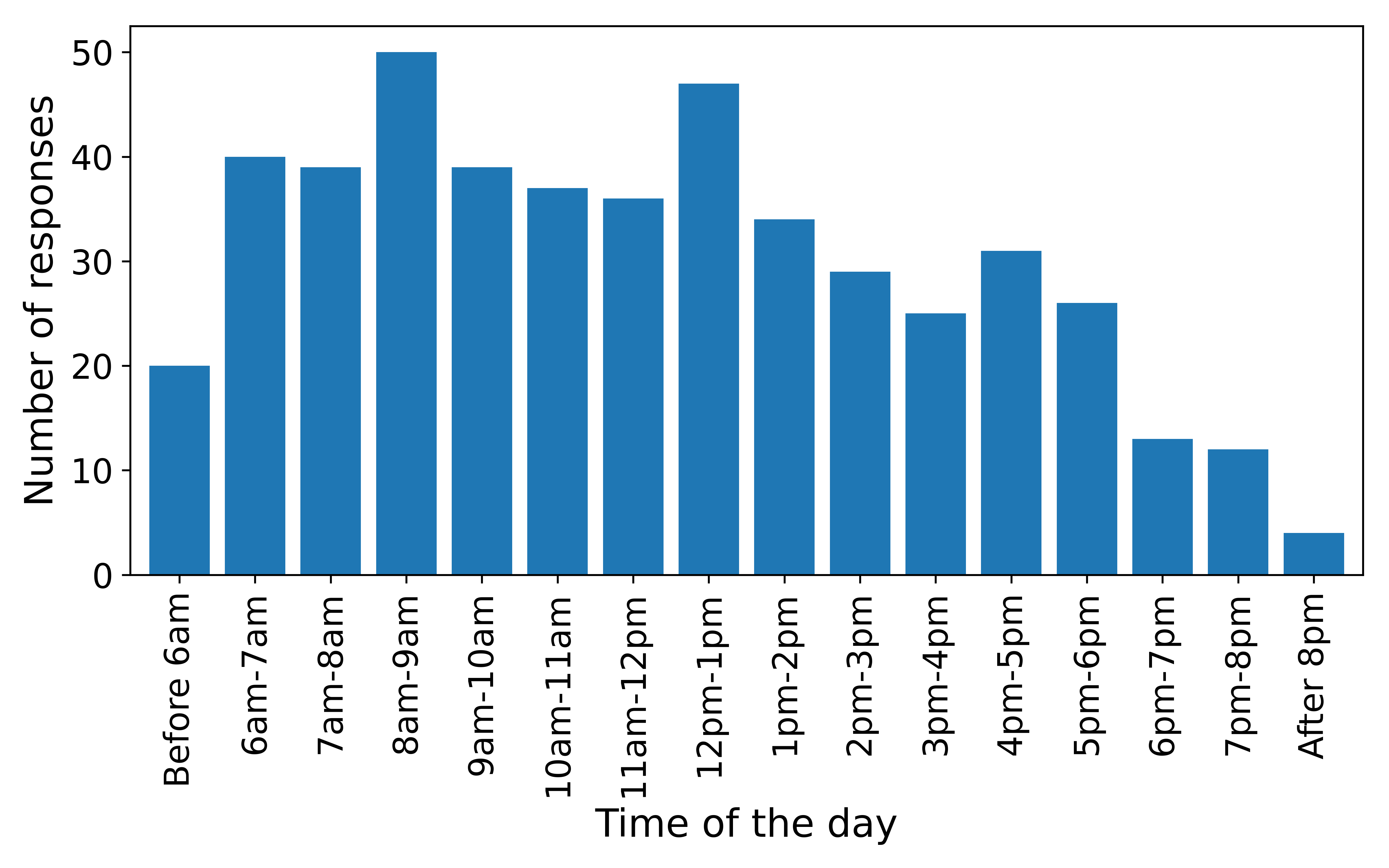}
    \caption{Self-reporting Times}
    \label{fig:resTime}
  \end{subfigure}

  \caption{Distribution of self-reported responses.}
  \label{fig:subfigures}
\end{figure}

\subsection{Response Distribution}

To better understand the collected data, we did a preliminary analysis of the features extracted from the blood volume signal and the responses given by the participants. First, we investigated the distribution of variables: the particular emotions, the cognitive load and the overall positive and negative affect.
We note that the \textit{cognitive load} (Fig. \ref{fig:cognitive_load}) and the overall \textit{positive affect} (Fig. \ref{fig:paffect}) are approximately normally distributed across the responses skewing slightly towards more positive scores, while the \textit{negative affect} is skewed towards the lower numbers (less negative responses) - Fig. \ref{fig:naffect}.  This trend is also true for the individual positive and negative emotions that add up to the overall affect.
% Participants most frequently reported their emotional state in the morning between 8-9 am and during the lunch break between 12-1 pm.
Interestingly,  frequently participants reported their emotional state prior to the first notification, indicating participants' prompt engagement upon waking up or arriving at the office around 8 am (See Fig. \ref{fig:resTime}).

\subsection{Relationship between HRV and self-reported states}

\begin{figure}[t]
    \centering
        \includegraphics[width=\textwidth]{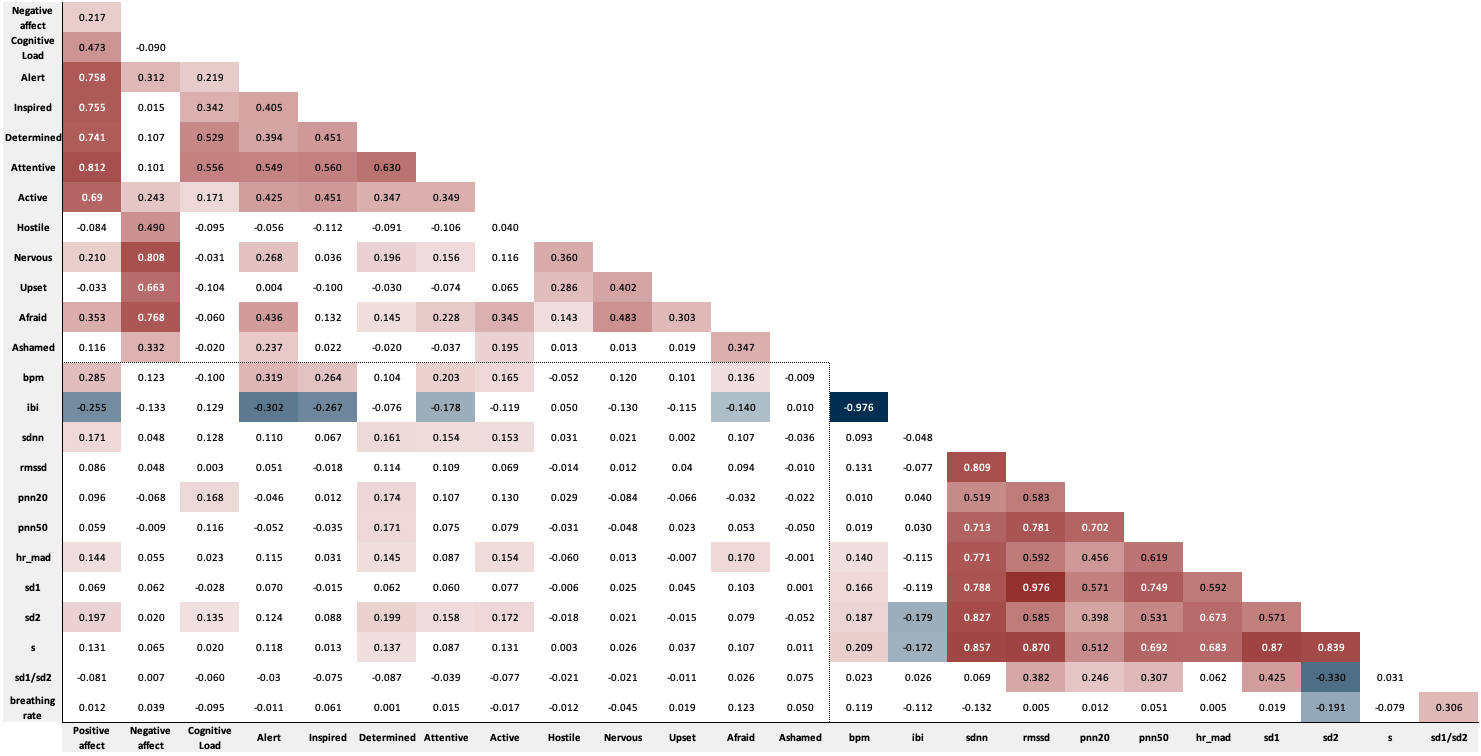}
    \caption{The figure shows the Pearson correlations computed for the self-reported emotional scores and the features extracted from the blood volume signal. The correlations that were found to be statistically significant (p < 0.05) are shown in colour, with red corresponding to the positive correlations and blue corresponding to the negative correlations. The upper left corner shows how the relationship between the self-reported emotions, including the cognitive load and the calculated overall positive and negative affect. The bottom right corner focuses on the relationships between the features extracted from the blood volume signal. The area within the dotted lines shows the correlations between the emotional scores and the signal obtained from the wearable device.}
    \label{fig:combined}
\end{figure}

Emotional states can influence blood volume and vice versa. During emotional arousal, such as excitement or fear, there is an increase in sympathetic nervous system activity, leading to vasoconstriction (narrowing of blood vessels) in certain areas of the body and vasodilation (widening of blood vessels) in others. This redistribution of blood volume is part of the body's physiological response to emotions. While it is challenging to directly infer specific emotional states from the PPG signal alone, certain patterns and features within the signal can provide insights into general physiological arousal or changes associated with emotions. To infer the relationship between the features extracted from the blood volume signal and the emotional states, we compute a matrix of Pearson's correlations as shown in Fig. \ref{fig:combined}.

\subsubsection{Correlations within the self-reported emotional scores}
We observe strong positive correlations between the \textit{positive affect} and its component emotions (between 0.69 and 0.81) and moderate to strong positive correlations between the \textit{negative affect} and its component emotions (0.33 to 0.81). There are also moderate correlations between individual positive emotions and individual negative emotions. Interestingly, some of the positive emotions are also correlated with the negative ones, specifically the emotions with high arousal. For example, \textit{alert} and \textit{active} are also correlated with the \textit{negative affect}, while \textit{nervous} and \textit{afraid} have positive correlations with the \textit{positive affect}. Additionally, moderate positive correlations are present between \textit{cognitive load} and the \textit{positive affect} and its component emotions (0.17 - 0.56).

\subsubsection{Correlations between the features extracted from the blood volume signal}
We observe many strong positive correlations between the signal features. Notably, the breathing rate has the fewest significant correlations with the other signal features. It is followed by the \textit{ibi}, which is negatively correlated with other signal features, including one very strong negative correlation with the \textit{bpm} (-0.98) and two weaker negative correlations between with the \textit{sd2} (-0.18) and \textit{s} (-0.17). Other negative correlations between signal features are observed between \textit{sd2} and the \textit{sd1/sd2} as well as \textit{sd2} and the \textit{breathing rate}.

\subsubsection{Correlations between the emotional scores and the signal features}
We did not observe any significant correlations between the \textit{negative affect} and the signal features with the exception of the component emotion of \textit{afraid}, which positively correlated with the \textit{bpm} (0.13) and the \textit{hr-mad} (0.17) and negatively correlated with the \textit{ibi}. In contrast, we identified several statistically significant correlations between the signal features and the \textit{cognitive load}, the \textit{positive affect} and the constituent positive emotions. The emotion of \textit{determined} is especially prominent with the biggest number of significant correlations with the signal features.

Notably, the \textit{ibi} has a moderate negative correlation with \textit{positive affect} (-0.26). Relaxation techniques and positive emotional states can promote parasympathetic dominance and increased HRV. This may lead to longer \textit{IBIs} and more variability between heartbeats. On the other hand, \textit{bpm}, \textit{sdnn}, \textit{hr-mad} and \textit{sd2} all have weak to moderate positive correlations with the \textit{positive affect} (0.14 - 0.29). 

Emotional arousal, such as during stress or excitement, often leads to increased sympathetic nervous system activity, resulting in an elevated heart rate and greater heart rate variability. \textit{BPM} tends to be higher than at baseline, reflecting the physiological response to emotional arousal, with higher \textit{hr-mad}. Simultaneously, positive emotions and engaging in emotional regulation strategies, such as deep breathing or mindfulness, can modulate the autonomic nervous system and impact heart rate variability. These techniques may promote increased parasympathetic activity, also leading to lower bpm and higher \textit{hr-mad}, \textit{sdnn} and \textit{sd2} values.

The state of positive relaxation can promote a decrease in heart rate. As a result, we would expect the \textit{bpm} to be lower during periods of relaxation or positive emotions. However, the relationship between \textit{bpm} and emotional states can vary among individuals due to individual differences. Factors such as fitness level, age, and overall cardiovascular health can influence baseline heart rate and its response to emotional stimuli. On top of that, our positive emotions mostly represent high arousal states, which could indicate that the higher \textit{bpm} is related to positive emotional arousal.

\subsection{Emotional valence classification}
\begin{table}[t]
\begin{tabular}{lcccc}
\hline
\multicolumn{1}{r}{\textbf{}}                                       & \textbf{Accuracy} & \textbf{F1} & \textbf{Precision} & \textbf{Recall} \\ \hline
\textbf{\begin{tabular}[c]{@{}l@{}}Positive\\ Affect\end{tabular}} & 0.648             & 0.645       & 0.646              & 0.648           \\
\textbf{Alert}  & 0.628 & 0.609 & 0.600 & 0.628 \\
\textbf{Afraid} & 0.621 & 0.741 & 0.947 & 0.621 \\
\textbf{Active} & 0.650 & 0.671 & 0.719 & 0.650 \\ \hline
\end{tabular}
\caption{Classification results for positive affect and three high arousal emotions as measured by accuracy and weighted F1 score, precision and recall. }
\label{tab:clasification_results}
\end{table}

In our investigation, we placed particular emphasis on examining the overall \textit{positive affect} due to its stronger correlations with signal features, as well as its more normal distribution among participants compared to the overall \textit{negative affect}.
Only 216 of the responses had corresponding HRV features, this might be due to missing signal which was not captured when participants used a mobile app, or due to poor signal quality which was prohibitive of reliable feature extraction.  Therefore, we simplified the \textit{positive affect} classification problem by transforming it into a binary classification task. The dataset was divided into two classes: responses with higher positive affect (minimum score of 17) and responses with lower positive affect (maximum score of 14). This division ensured a balanced number of responses in each class, excluding neutral scores. High positive affect was assigned the label 1, while low positive affect was assigned the label -1. For classification, we utilized the Multinomial Naive Bayes model implemented in the scikit-learn library \cite{scikit-learn}. The number of responses between participants strongly varies. To reduce the risk of overfitting to individuals and ensure diversity, for each participant, we utilized the first two-thirds of their responses for training, reserving the remaining responses for testing purposes.  
Our model achieved an accuracy rate of 65\% in differentiating between high and low positive emotional states (see Table \ref{tab:clasification_results}). 

We also assessed the classification performance when differentiating between high ($\geq$ 4) and low ($\leq$ 2)  ratings for specific emotions for which we obtain a sufficient amount of highly and lowly scored data namely \textit{Alert}, \textit{Afraid}, and \textit{Active}. It is important to note that these emotions represent high arousal states but exhibit varying levels of valence. Interestingly, we achieved a similar level of performance (62-65\%) in these classification tasks.

\section{Discussion and conclusions}
\label{sec:discussion}
Wearable devices that can detect emotional states have the potential to revolutionize the field of mental health by providing individuals with greater insight into their emotional patterns and helping healthcare providers develop more personalized and effective mobile health interventions. The findings from our small-scale feasibility study shed light on the potential use of consumer-grade wearables for the automatic detection of individuals' emotional valence. Through our initial analysis, we observed correlations between certain HRV features, such as \textit{ibi}, \textit{sdnn} as well as \textit{sd2}, and \textit{positive affect}. This implies that these features can provide valuable information about user valence.

Despite the limited amount of training data available for the machine learning models, the preliminary results for positive affect classification showed promise. This suggests that consumer-grade wearables have the capability to effectively capture emotional valence. Furthermore, we also observed encouraging initial classification results for some high-arousal emotions.
However, it is important to acknowledge the limitations of our study (See Section\ref{sec:limitation}), including the small sample size and the need for further research and validation using larger datasets. These additional investigations will help strengthen the validity and generalizability of our findings.

In conclusion, our study highlights the potential of consumer-grade wearables in capturing and detecting emotional valence. The correlations between physiological signal and positive affect provide insights for future research and development in the field of mobile mental health interventions but also open discussion on what should or could be done with this information (See Section \ref{sec:ethics}).

\subsection{Ethical Considerations in MobileHCI Context}
\label{sec:ethics}

Collecting and analyzing personal emotional data from wearable devices raises significant privacy concerns and the potential for misuse. However, from a machine learning perspective, there are approaches that can help address these privacy concerns, such as federated learning setups and differential privacy techniques. Despite these measures, there remains a risk that accurately detected emotions could be exploited to manipulate individuals' behavior or exploit their vulnerabilities, highlighting the ethical considerations surrounding emotion detection.

On a positive note, emotion detection has the potential to contribute to the development of personalized and context-aware mobile applications. By understanding users' emotional states, systems can adapt and tailor their responses to better meet their needs. This can result in improved user experiences, enhanced productivity, and more effective interventions in fields like mental health and well-being. Moreover, emotion detection can foster the creation of more inclusive and empathetic technologies. By recognizing users' emotions, systems can respond with empathy, understanding, and appropriate support. This has the potential to enhance human-computer interactions and cultivate positive emotional experiences.

\subsection{Limitations and Future Work}
\label{sec:limitation}

During our initial data exploration, we observed a tendency for participants to report lower ratings for negative valence emotions such as nervousness and feeling upset, with rare occurrences of high scores.  To ascertain the validity of the dataset and identify potential sources of bias, a more extensive analysis is required. The skewed distribution of reported scores suggests that participants may generally experience positive moods or be less inclined to report negative emotions. Consequently, we opted to reframe the valence classification problem by differentiating between positive and non-positive emotional states, instead of positive and negative states. These observations of reporting bias have important implications for the design of future well-being studies that utilize mobile applications for self-reporting and the development of active learning data annotation pipelines for machine learning-based emotion classification. One potential strategy to encourage more annotations is to phrase ratings solely on a positive scale. Additionally, the refusal to respond to certain emotions may serve as an indirect indicator of a negative emotional state.

To minimize the reporting effort required by study participants, we employed a short PANAS questionnaire consisting of only 10 questions. However, it is worth noting that PANAS primarily captures positive and negative emotions that predominantly lie on the high arousal spectrum. Emotions that are very low in arousal, such as relaxation, are not assessed. Therefore, in our analysis, we focused solely on the valence dimension of emotions and did not consider emotions in a 2D space where arousal is assigned to the first dimension and valence to the second dimension, as proposed by Russell \cite{russell1979affective}. To address the arousal dimension in future studies, it would be necessary to include questions that capture low arousal positive emotions.

In order to validate our initial results and potentially train a detector capable of classifying all 10 emotions, we plan to collect data from a larger number of participants. Currently, we have encountered challenges with missing HRV features associated with a significant number of PANAS responses.  When participants responded using their phones instead of the watch, the PGG signal was missing. The connectivity issues might have been another cause of missing signal data. Some samples had PPG signal captured but the quality was insufficient to extract HRV features. Therefore the data collection protocol needs to be refined and data from a larger number of participants collected. 
This expanded dataset will provide a more robust foundation for our analysis and allow for further investigation into emotion classification.

% UNFORTUNATELY THIS NEEDS TO BE COMMENTED DURING ANONYMOUS REVIEW PROCESS, UN-COMMENT ONCE ACCEPTED
\begin{acks}
    This publication is partly supported by the European Union’s Horizon 2020 research and innovation programme under grant agreement Sano No. 857533 and the International Research Agendas programme of the Foundation for Polish Science, co-financed by the European Union under the European Regional Development Fund.
\end{acks}

% %%
% %% The next two lines define the bibliography style to be used, and
% %% the bibliography file.
\bibliographystyle{ACM-Reference-Format}
\bibliography{ref}


\begin{thebibliography}{16}

%%% the Web address.

\ifx \showCODEN    \undefined \def \showCODEN     #1{\unskip}     \fi
\ifx \showDOI      \undefined \def \showDOI       #1{#1}\fi
\ifx \showISBNx    \undefined \def \showISBNx     #1{\unskip}     \fi
\ifx \showISBNxiii \undefined \def \showISBNxiii  #1{\unskip}     \fi
\ifx \showISSN     \undefined \def \showISSN      #1{\unskip}     \fi
\ifx \showLCCN     \undefined \def \showLCCN      #1{\unskip}     \fi
\ifx \shownote     \undefined \def \shownote      #1{#1}          \fi
\ifx \showarticletitle \undefined \def \showarticletitle #1{#1}   \fi
\ifx \showURL      \undefined \def \showURL       {\relax}        \fi
% invisible to TeX
\providecommand\bibfield[2]{#2}
\providecommand\bibinfo[2]{#2}
\providecommand\natexlab[1]{#1}
\providecommand\showeprint[2][]{arXiv:#2}

\bibitem[Can et~al\mbox{.}(2019a)]%
        {can2019stress}
\bibfield{author}{\bibinfo{person}{Yekta~Said Can}, \bibinfo{person}{Bert
  Arnrich}, {and} \bibinfo{person}{Cem Ersoy}.}
  \bibinfo{year}{2019}\natexlab{a}.
\newblock \showarticletitle{Stress detection in daily life scenarios using
  smart phones and wearable sensors: A survey}.
\newblock \bibinfo{journal}{\emph{Journal of biomedical informatics}}
  \bibinfo{volume}{92} (\bibinfo{year}{2019}), \bibinfo{pages}{103139}.
\newblock


\bibitem[Can et~al\mbox{.}(2019b)]%
        {can2019continuous}
\bibfield{author}{\bibinfo{person}{Yekta~Said Can}, \bibinfo{person}{Niaz
  Chalabianloo}, \bibinfo{person}{Deniz Ekiz}, {and} \bibinfo{person}{Cem
  Ersoy}.} \bibinfo{year}{2019}\natexlab{b}.
\newblock \showarticletitle{Continuous stress detection using wearable sensors
  in real life: Algorithmic programming contest case study}.
\newblock \bibinfo{journal}{\emph{Sensors}} \bibinfo{volume}{19},
  \bibinfo{number}{8} (\bibinfo{year}{2019}), \bibinfo{pages}{1849}.
\newblock


\bibitem[Grzeszczyk et~al\mbox{.}(2023)]%
        {grzeszczykcogwear}
\bibfield{author}{\bibinfo{person}{Michal~K. Grzeszczyk},
  \bibinfo{person}{Rosmary Blanco}, \bibinfo{person}{Paulina Adamczyk},
  \bibinfo{person}{Maciej Kuś}, \bibinfo{person}{Sylwia Marek},
  \bibinfo{person}{Ryszard Pręcikowski}, {and} \bibinfo{person}{Aneta
  Lisowska}.} \bibinfo{year}{2023}\natexlab{}.
\newblock \showarticletitle{CogWear: Can we detect cognitive effort with
  consumer-grade wearables?' (version 1.0.0)}.
\newblock \bibinfo{journal}{\emph{PhysioNet}} (\bibinfo{year}{2023}).
\newblock


\bibitem[Markova et~al\mbox{.}(2019)]%
        {markova2019clas}
\bibfield{author}{\bibinfo{person}{Valentina Markova}, \bibinfo{person}{Todor
  Ganchev}, {and} \bibinfo{person}{Kalin Kalinkov}.}
  \bibinfo{year}{2019}\natexlab{}.
\newblock \showarticletitle{Clas: A database for cognitive load, affect and
  stress recognition}. In \bibinfo{booktitle}{\emph{2019 International
  Conference on Biomedical Innovations and Applications (BIA)}}. IEEE,
  \bibinfo{pages}{1--4}.
\newblock


\bibitem[Oakley-Girvan and \etal(2021)]%
        {oakley2021works}
\bibfield{author}{\bibinfo{person}{Ingrid Oakley-Girvan} {and}
  \bibinfo{person}{\etal}.} \bibinfo{year}{2021}\natexlab{}.
\newblock \showarticletitle{What Works Best to Engage Participants in Mobile
  App Interventions and e-Health: A Scoping Review}.
\newblock \bibinfo{journal}{\emph{Telemedicine and e-Health}}
  (\bibinfo{year}{2021}).
\newblock


\bibitem[Park et~al\mbox{.}(2020)]%
        {park2020k}
\bibfield{author}{\bibinfo{person}{Cheul~Young Park}, \bibinfo{person}{Narae
  Cha}, \bibinfo{person}{Soowon Kang}, \bibinfo{person}{Auk Kim},
  \bibinfo{person}{Ahsan~Habib Khandoker}, \bibinfo{person}{Leontios
  Hadjileontiadis}, \bibinfo{person}{Alice Oh}, \bibinfo{person}{Yong Jeong},
  {and} \bibinfo{person}{Uichin Lee}.} \bibinfo{year}{2020}\natexlab{}.
\newblock \showarticletitle{K-EmoCon, a multimodal sensor dataset for
  continuous emotion recognition in naturalistic conversations}.
\newblock \bibinfo{journal}{\emph{Scientific Data}} \bibinfo{volume}{7},
  \bibinfo{number}{1} (\bibinfo{year}{2020}), \bibinfo{pages}{293}.
\newblock


\bibitem[Patel and Saunders(2018)]%
        {patel2018apps}
\bibfield{author}{\bibinfo{person}{Sunil Patel} {and} \bibinfo{person}{Kate~EA
  Saunders}.} \bibinfo{year}{2018}\natexlab{}.
\newblock \showarticletitle{Apps and wearables in the monitoring of mental
  health disorders}.
\newblock \bibinfo{journal}{\emph{British Journal of Hospital Medicine}}
  \bibinfo{volume}{79}, \bibinfo{number}{12} (\bibinfo{year}{2018}),
  \bibinfo{pages}{672--675}.
\newblock


\bibitem[Patoz et~al\mbox{.}(2021)]%
        {patoz2021patient}
\bibfield{author}{\bibinfo{person}{Marie-Camille Patoz}, \bibinfo{person}{Diego
  Hidalgo-Mazzei}, \bibinfo{person}{Olivier Blanc}, \bibinfo{person}{Norma
  Verdolini}, \bibinfo{person}{Isabella Pacchiarotti}, \bibinfo{person}{Andrea
  Murru}, \bibinfo{person}{Laurent Zukerwar}, \bibinfo{person}{Eduard Vieta},
  \bibinfo{person}{Pierre-Michel Llorca}, {and} \bibinfo{person}{Ludovic
  Samalin}.} \bibinfo{year}{2021}\natexlab{}.
\newblock \showarticletitle{Patient and physician perspectives of a smartphone
  application for depression: a qualitative study}.
\newblock \bibinfo{journal}{\emph{BMC psychiatry}} \bibinfo{volume}{21},
  \bibinfo{number}{1} (\bibinfo{year}{2021}), \bibinfo{pages}{1--12}.
\newblock


\bibitem[Peake et~al\mbox{.}(2018)]%
        {peake2018critical}
\bibfield{author}{\bibinfo{person}{Jonathan~M Peake}, \bibinfo{person}{Graham
  Kerr}, {and} \bibinfo{person}{John~P Sullivan}.}
  \bibinfo{year}{2018}\natexlab{}.
\newblock \showarticletitle{A critical review of consumer wearables, mobile
  applications, and equipment for providing biofeedback, monitoring stress, and
  sleep in physically active populations}.
\newblock \bibinfo{journal}{\emph{Frontiers in physiology}}
  \bibinfo{volume}{9} (\bibinfo{year}{2018}), \bibinfo{pages}{743}.
\newblock


\bibitem[Pedregosa et~al\mbox{.}(2011)]%
        {scikit-learn}
\bibfield{author}{\bibinfo{person}{F. Pedregosa}, \bibinfo{person}{G.
  Varoquaux}, \bibinfo{person}{A. Gramfort}, \bibinfo{person}{V. Michel},
  \bibinfo{person}{B. Thirion}, \bibinfo{person}{O. Grisel},
  \bibinfo{person}{M. Blondel}, \bibinfo{person}{P. Prettenhofer},
  \bibinfo{person}{R. Weiss}, \bibinfo{person}{V. Dubourg}, \bibinfo{person}{J.
  Vanderplas}, \bibinfo{person}{A. Passos}, \bibinfo{person}{D. Cournapeau},
  \bibinfo{person}{M. Brucher}, \bibinfo{person}{M. Perrot}, {and}
  \bibinfo{person}{E. Duchesnay}.} \bibinfo{year}{2011}\natexlab{}.
\newblock \showarticletitle{Scikit-learn: Machine Learning in {P}ython}.
\newblock \bibinfo{journal}{\emph{Journal of Machine Learning Research}}
  \bibinfo{volume}{12} (\bibinfo{year}{2011}), \bibinfo{pages}{2825--2830}.
\newblock


\bibitem[Picard(2016)]%
        {picard2016automating}
\bibfield{author}{\bibinfo{person}{Rosalind~W Picard}.}
  \bibinfo{year}{2016}\natexlab{}.
\newblock \showarticletitle{Automating the recognition of stress and emotion:
  From lab to real-world impact}.
\newblock \bibinfo{journal}{\emph{IEEE MultiMedia}} \bibinfo{volume}{23},
  \bibinfo{number}{3} (\bibinfo{year}{2016}), \bibinfo{pages}{3--7}.
\newblock


\bibitem[Russell(1979)]%
        {russell1979affective}
\bibfield{author}{\bibinfo{person}{James~A Russell}.}
  \bibinfo{year}{1979}\natexlab{}.
\newblock \showarticletitle{Affective space is bipolar.}
\newblock \bibinfo{journal}{\emph{Journal of personality and social
  psychology}} \bibinfo{volume}{37}, \bibinfo{number}{3}
  (\bibinfo{year}{1979}), \bibinfo{pages}{345}.
\newblock


\bibitem[Saganowski et~al\mbox{.}(2020)]%
        {saganowski2020consumer}
\bibfield{author}{\bibinfo{person}{Stanislaw Saganowski},
  \bibinfo{person}{Przemyslaw Kazienko}, \bibinfo{person}{Maciej Dziezyc},
  \bibinfo{person}{Patrycja Jakimow}, \bibinfo{person}{Joanna Komoszynska},
  \bibinfo{person}{Weronika Michalska}, \bibinfo{person}{Anna Dutkowiak},
  \bibinfo{person}{Adam Polak}, \bibinfo{person}{Adam Dziadek}, {and}
  \bibinfo{person}{Michal Ujma}.} \bibinfo{year}{2020}\natexlab{}.
\newblock \showarticletitle{Consumer wearables and affective computing for
  wellbeing support}. In \bibinfo{booktitle}{\emph{MobiQuitous 2020-17th EAI
  International Conference on Mobile and Ubiquitous Systems: Computing,
  Networking and Services}}. \bibinfo{pages}{482--487}.
\newblock


\bibitem[Schmidt et~al\mbox{.}(2018)]%
        {schmidt2018introducing}
\bibfield{author}{\bibinfo{person}{Philip Schmidt}, \bibinfo{person}{Attila
  Reiss}, \bibinfo{person}{Robert Duerichen}, \bibinfo{person}{Claus
  Marberger}, {and} \bibinfo{person}{Kristof Van~Laerhoven}.}
  \bibinfo{year}{2018}\natexlab{}.
\newblock \showarticletitle{Introducing wesad, a multimodal dataset for
  wearable stress and affect detection}. In
  \bibinfo{booktitle}{\emph{Proceedings of the 20th ACM international
  conference on multimodal interaction}}. \bibinfo{pages}{400--408}.
\newblock


\bibitem[Thompson(2007)]%
        {thompson2007development}
\bibfield{author}{\bibinfo{person}{Edmund~R Thompson}.}
  \bibinfo{year}{2007}\natexlab{}.
\newblock \showarticletitle{Development and validation of an internationally
  reliable short-form of the positive and negative affect schedule (PANAS)}.
\newblock \bibinfo{journal}{\emph{Journal of cross-cultural psychology}}
  \bibinfo{volume}{38}, \bibinfo{number}{2} (\bibinfo{year}{2007}),
  \bibinfo{pages}{227--242}.
\newblock


\bibitem[Van~Gent et~al\mbox{.}(2018)]%
        {van2018heart}
\bibfield{author}{\bibinfo{person}{Paul Van~Gent}, \bibinfo{person}{Haneen
  Farah}, \bibinfo{person}{Nicole Nes}, {and} \bibinfo{person}{Bart van Arem}.}
  \bibinfo{year}{2018}\natexlab{}.
\newblock \showarticletitle{Heart rate analysis for human factors: Development
  and validation of an open source toolkit for noisy naturalistic heart rate
  data}. In \bibinfo{booktitle}{\emph{Proceedings of the 6th HUMANIST
  Conference}}. \bibinfo{pages}{173--178}.
\newblock


\end{thebibliography}

\end{document}